\documentstyle[pra,aps,twocolumn,psfig,floats,amssymb]{revtex}

\def\ba{ \begin{array} }
\def\ea{ \end{array} }
\def\be{ \begin{equation}\label }
\def\ee{ \end{equation} }
\def\bea{ \begin{eqnarray} }
\def\eea{ \end{eqnarray} }
\def\bml{ \begin{mathletters} \label }
\def\eml{ \end{mathletters} }

\def\emla{ \eea \eml }

\def\bM{\left[ \matrix}
\def\eM{ \right]}

\def\H{{\bf H}}
\def\a{\alpha}
\def\b{\beta}
\def\d{\delta}

\def\h{h}
\def\P{{\cal D}}
\def\&{\!&\!}
\def\={\!=\!}
\def\AT{\varphi_T}  
\def\dark{\varphi_D}  

\def\State#1{\varphi_{#1}}
\def\state#1{\psi_{#1}}
\def\S{{\cal S}}
\def\Sa{\S_{\a^2}}
\def\Sb{\S_{\b^2}}
\def\Sab{\S_{\a\b}}
\def\San{\Sa^{(n)}}
\def\Sbn{\Sb^{(n)}}
\def\Sabn{\Sab^{(n)}}

\def\EV{\lambda}
\def\OmP{\Omega_P}
\def\OmS{\Omega_S}
\def\Om0{\Omega_0}
\def\c{{\bf c}}
\def\Om{\Omega}
\def\Omeff{\Om_{\rm eff}}
\def\Deff{\Delta_{\rm eff}}
\def\LZ{\xi}
\def\text{\mbox} 

\begin{document}
\draft
\wideabs{
\title{Adiabatic population transfer via multiple intermediate states
}
\author{N. V. Vitanov$^{1,\dagger}$ and S. Stenholm$^{2}$}
\address{
$^1$Helsinki Institute of Physics, P.O.Box 9,
 FIN-00014 Helsingin yliopisto, Finland\\
$^2$Department of Physics, Royal Institute of Technology (KTH),
 Lindstedtv. 24, SE-10044 Stockholm, Sweden \\
$^\dagger$electronic address: vitanov@rock.helsinki.fi
}
\date{\today }
\maketitle
\begin{abstract}
This paper discusses a generalization of stimulated Raman adiabatic
passage (STIRAP) in which the single intermediate state
is replaced by $N$ intermediate states.
Each of these states is connected to the initial state $\state{i}$ with a
coupling proportional to the pump pulse and to the final state $\state{f}$
with a coupling proportional to the Stokes pulse,
thus forming a parallel multi-$\Lambda$ system.
It is shown that the dark (trapped) state exists only when
the ratio between each pump coupling and the respective Stokes coupling
is the same for all intermediate states.
We derive the conditions for existence of a more general adiabatic-transfer
state which includes transient contributions from the intermediate
states but still transfers the population from state $\state{i}$ to
state $\state{f}$ in the adiabatic limit.
We present various numerical examples for success and failure of
multi-$\Lambda$ STIRAP which illustrate the analytic predictions.
Our results suggest that in the general case of arbitrary couplings,
it is most appropriate to tune the pump and Stokes lasers either
just below or just above all intermediate states.
\end{abstract}
\pacs{32.80.Bx, 33.80.Be, 42.50.-p}
} 


\section{Introduction}

\label{Sec-introduction}

Stimulated Raman adiabatic passage (STIRAP)
\cite{Gaubatz88,Kuklinski89,Gaubatz90}
is an established technique for efficient population transfer
in three-state systems in $\Lambda$ or ladder configurations.
In the original STIRAP, the population is transferred adiabatically
from an initial state $\state{i}$ to a final target state $\state{f}$
via an intermediate state $\state{int}$
by means of two partly overlapping laser pulses,
a pump pulse $\OmP(t)$ linking states $\state{i}$ and $\state{int}$,
and a Stokes pulse $\OmS(t)$ linking states $\state{int}$ and $\state{f}$.
By applying the Stokes pulse before the pump pulse
(counterintuitive pulse order)
and maintaining adiabatic-evolution conditions,
one ensures population transfer from the initial state into the final
state, with negligible population in the intermediate state at any time.
This is so because the transfer is realized via an adiabatic state
$\dark(t)$ (instantaneous eigenstate of the Hamiltonian)
--- the so-called dark state --- which is
a linear superposition of states $\state{i}$ and $\state{f}$ only.
In the ideal limit, unit transfer efficiency is guaranteed
and the process is robust against moderate changes in the laser parameters.
Various aspects of STIRAP have been studied in detail theoretically
and experimentally \cite{Bergmann98}.
Among them are
the effects of intermediate-state detuning \cite{Vitanov97D}
and loss rate \cite{Vitanov97G},
two-photon detuning \cite{Danileiko94,Romanenko97,Fewell97},
nonadiabatic effects \cite{Laine96,Vitanov96,Drese98},
multiple intermediate and final states
\cite{Coulston92,Shore95,Martin95,Martin96}.
The success of STIRAP has encouraged its extensions in various
directions, such as population transfer in chainwise connected
multistate systems
\cite{Shore91,Marte91,Oreg92,Smith92,Pillet93,Valentin94,%
Malinovsky97,Theuer98,Vitanov98multi,Vitanov98tune}
and population transfer via continuum
\cite{Carroll92,Carroll93,Carroll95,Carroll96,Nakajima94,%
Yatsenko97,Vitanov97c,Paspalakis97,Unanyan98}.

In the present paper, we examine the possibilities to achieve complete
adiabatic population transfer in the case when the single intermediate
state in STIRAP is replaced by $N$ states, each of which is coupled to
the initial state $\state{i}$ with a coupling proportional to the pump
field and to the final state $\state{f}$ with a coupling proportional
to the Stokes field, thus forming a parallel multi-$\Lambda$ system.
In the first place, our work is motivated by the possibility that
appreciable single-photon couplings to more than one intermediate states
can exist in a {\it realistic physical situation},
for example, when the pump and Stokes lasers are tuned to a highly
excited state in an atom,
or in the case of population transfer in molecules.
Such couplings may be present because,
while very sensitive to the two-photon resonance
\cite{Danileiko94,Romanenko97},
STIRAP is relatively insensitive to the single-photon detuning
from the intermediate state \cite{Vitanov97D},
the detuning tolerance range being proportional
to the squared pump and Stokes Rabi frequencies.
It is therefore important to know if STIRAP-like transfer can
take place in such systems.
A second example for multi-$\Lambda$ systems can be found in population
transfer in {\it multistate chains}.
It has been shown that when the couplings between the intermediate
states are constant the multistate chain is mathematically equivalent to
a multi-$\Lambda$ system, in which the initial state $\state{i}$
is coupled simultaneously to $N-2$ dressed states which are in turn
coupled to the final state $\state{f}$ \cite{Vitanov98tune}.
It has been demonstrated that in certain domains of interaction
parameters (Rabi frequencies and detunings), adiabatic population
transfer in these multistate chains (respectively, in the equivalent
multi-$\Lambda$ systems) can take place,
while in other domains it cannot.
A third motivation for the present work is
{\it population transfer via continuum}
\cite{Carroll92,Carroll93,Carroll95,Carroll96,Nakajima94,%
Yatsenko97,Vitanov97c,Paspalakis97,Unanyan98}.
In their pioneering work on this process \cite{Carroll92}, Carroll and Hioe
have replaced the single descrete intermediate state in STIRAP
by a quasicontinuum consisting of infinite number of equidistant
discrete states with energies going from $-\infty$ to $+\infty$.
Moreover, each of these states was coupled with the same coupling $\OmP(t)$
to the initial state $\state{i}$ and with the same coupling $\OmS(t)$
to the final state $\state{f}$.
Under these conditions Carroll and Hioe have shown that complete
population transfer is achieved in the adiabatic limit with
counterintuitively ordered pulses.
It has been shown later \cite{Nakajima94} that the Carroll-Hioe's
quasicontinuum is too simplified and symmetric, and that a real
continuum has properties, such as nonzero Fano parameter,
which prevent complete population transfer.
It is interesting to find out which of the simplifying assumptions
in the Carroll-Hioe model
(equidistant states,
going to infinity both upwards and downwards,
equal pump couplings and equal Stokes couplings)
makes the dark state in the original STIRAP remain as
a zero-eigenvalue eigenstate of the multi-$\Lambda$ Hamiltonian.
In this paper, we consider the general asymmetric case of unequal
couplings and unevenly distributed, finite number intermediate states.
Besides the Carroll-Hioe model \cite{Carroll92},
our paper generalizes the results of Coulston and Bergmann
\cite{Coulston92} who were the first to consider the effects of
multiple intermediate states in the simplest case of $N=2$
states and equal couplings $\OmP(t)$ to state $\state{i}$
and equal couplings $\OmS(t)$ to state $\state{f}$.
We shall find the conditions for the existence of the dark state
in multi-$\Lambda$ systems as well as the conditions for the existence
of a more general adiabatic-transfer state, which still links states
$\state{i}$ and $\state{f}$ adiabatically but is allowed to contain
transient contributions from the intermediate states.

Our paper is organized as follows.
In Sec.~\ref{Sec-definition} we present the basic equations
and definitions and review the standard three-state STIRAP.
In Sec.~\ref{Sec-off} we discuss the case when all intermediate states
are off single-photon resonance.
In Sec.~\ref{Sec-on} we consider the case when one of the intermediate
states is on single-photon resonance and in Sec.~\ref{Sec-degenerate}
the case of degenerate resonant states.
In Sec.~\ref{Sec-AE} we use the adiabatic-elimination approximation to
gain further insight of the process.
Finally, in Sec.~\ref{Sec-conclusion} we summarize the conclusions.


\section{Basic equations and definitions}

\label{Sec-definition}

\subsection{Basic STIRAP}

The probability amplitudes of the three states in STIRAP
satisfy the Schrödinger equation ($\hbar = 1$),
\be{SEq}
 i \frac d{dt}{\c}(t) = \H(t) {\c}(t),
\ee
where $\c(t)=[c_i(t),c_{int}(t),c_f(t)]^T$.
In the rotating-wave approximation \cite{Shore90},
the Hamiltonian is given by
\be{H3}
\H(t)=
\bM{
0       & \OmP(t) & 0 \cr
\OmP(t) & \Delta  & \OmS(t) \cr
0       & \OmS(t) & 0
}\eM.
\ee
The time-varying Rabi frequencies $\OmP(t)$ and $\OmS(t)$
are given by products of the corresponding transition dipole moments
and electric-field amplitudes.
States $\state{i}$ and $\state{f}$ are assumed to be on two-photon
resonance, while the intermediate state $\state{int}$ may be off
single-photon resonance by a detuning $\Delta$.
The system is initially in state $\state{i}$,
$c_i(-\infty)=1,\ c_{int}(-\infty)=c_f(-\infty)=0$,
and the quantities of interest are the populations, and partucularly the
population of state $\state{f}$ at $t\rightarrow +\infty$,
$P_f(\infty)=|c_f(\infty)|^2$.

Throughout this paper, we assume that the two pulses are ordered
counterintuitively, i.e., the Stokes pulse precedes the pump pulse,
\be{CIorder}
\lim_{t\rightarrow -\infty} \frac{\OmP(t)}{\OmS(t)} = 0,\qquad
\lim_{t\rightarrow +\infty} \frac{\OmS(t)}{\OmP(t)} = 0,
\ee
but we do not impose any restrictions on the particular time dependences
in our analysis.
In the numerical examples, we assume Gaussian shapes,
\be{shapes}
\OmP(t) = \Om_0 e^{-(t-\tau)^2/T^2},\ \ 
\OmS(t) = \Om_0 e^{-(t+\tau)^2/T^2},
\ee
where $T$ is the pulse width, $2\tau$ is the time delay between the pulses,
and we take $\tau=0.5T$ everywhere.
Furthermore, we choose the peak Rabi frequency $\Om_0$
to define the frequency and time scales.

The essence of STIRAP is explained in terms of the so-called
dark (or trapped) state $\dark(t)$, which is a
zero-eigenvalue eigenstate of $\H(t)$,
\be{dark}
\dark(t) = \frac{\OmS(t)}{\Om(t)}\state{i} - \frac{\OmP(t)}{\Om(t)}\state{f},
\ee
where $\Om(t)$ is the mean-square Rabi frequency,
\be{Omega}
\Om(t)=\sqrt{\OmS^2(t)+\OmP^2(t)}.
\ee
For counterintuitively ordered pulses, Eq.~(\ref{CIorder}),
we have $\dark(-\infty)=\state{i}$ and $\dark(+\infty)=-\state{f}$ and hence,
the dark state connects adiabatically states $\state{i}$ and $\state{f}$.
By maintaining adiabatic evolution
(a condition which amounts to requiring that the pulse width $T$ is large
or that the pulse areas are much larger than $\pi$), one can force the
system to remain in the dark state and achieve complete population
transfer from $\state{i}$ to $\state{f}$.
Moreover, since $\dark(t)$ does not involve the intermediate state
$\state{int}$, the latter is not populated in the adiabatic limit,
even transiently, and hence, its properties, including decay,
do not affect the transfer efficiency.


\subsection{Multi-$\Lambda$ STIRAP}

\subsubsection{The system}


\begin{figure}
\centerline{\psfig{width=80mm,file=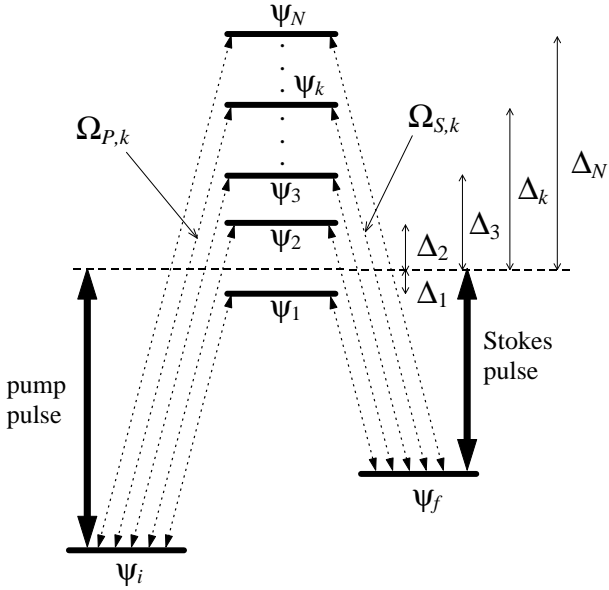}}
\vspace*{3mm}
\caption{
The multi-$\Lambda$ system studied in this paper.
The system is initially in state $\state{i}$
and the objective is to transfer it to state $\state{f}$
via one or more intermediate states $\state{1},\state{2},\ldots,\state{N}$
by means of two delayed laser pulses,
the pump pulse $\OmP(t)$ and the Stokes pulse $\OmS(t)$,
ordered counterintuitively (the Stokes pulse coming first).
Each intermediate state is coupled to state $\state{i}$ with a coupling
$\Om_{P,k}(t)$ proportional to $\OmP(t)$ and to state $\state{f}$
with a coupling $\Om_{S,k}(t)$ proportional to $\OmS(t)$.
Each intermediate state $\State{k}$ is detuned from single-photon
resonance with a detuning $\Delta_k$ while states $\state{i}$
and $\state{f}$ are on two-photon resonance.
}
\label{Fig-system}
\end{figure}


In the multi-$\Lambda$ generalization of STIRAP the single
intermediate state is replaced by $N$ intermediate states
$\state{1},\state{2},\ldots,\state{N}$, each of which is coupled to both
states $\state{i}$ and $\state{f}$, as shown in Fig.~\ref{Fig-system}.
The system is again assumed to be initially in state $\state{i}$,
\bml{initial}\bea
&&c_i(-\infty)=1,\\
&&c_f(-\infty)=c_1(-\infty )=\ldots=c_N(-\infty )=0,
\emla
and the objective is to transfer the population to state $\state{f}$.
The column vector of the probability amplitudes in 
the Schrödinger equation (\ref{SEq}) is given by
$\c(t)=[c_i(t),c_1(t),\ldots,c_N(t),c_f(t)]^T$
and the Hamiltonian reads
\be{H}
\H =\!
\left[\ba{ccccccc}\!\!
0\& \Om_{P,1}\& \Om_{P,2}\& \cdots\& \Om_{P,N-1}\& \Om_{P,N}\& 0\\
\Om_{P,1} \& \Delta_1 \&  0 \& \cdots \& 0 \& 0 \& \Om_{S,1} \\
\Om_{P,2} \& 0 \&  \Delta_2 \& \cdots \& 0 \& 0 \& \Om_{S,2} \\
\vdots \& \vdots \& \vdots \& \ddots \& \vdots \& \vdots \& \vdots \\
\Om_{P,N-1} \& 0 \&  0 \& \cdots \& \Delta_{N-1} \& 0 \& \Om_{S,N-1} \\
\Om_{P,N} \& 0 \& 0 \& \cdots \& 0 \& \Delta_N \& \Om_{S,N} \\
0\& \Om_{S,1}\& \Om_{S,2}\& \cdots\& \Om_{S,N-1}\& \Om_{S,N}\& 0
\ea \!\! \right]\!\!.
\ee
States $\state{i}$ and $\state{f}$ are again assumed to be on two-photon
resonance while each intermediate state $\state{k}$ may be off single-photon
resonance by a detuning $\Delta_k$.
The couplings $\Om_{P,k}(t)$ of the intermediate states $\state{k}$ to the
initial state $\state{i}$ are proportional to the pump field,
while the couplings $\Om_{S,k}(t)$ of the intermediate states
to the final state $\state{f}$ are proportional to the Stokes field,
\be{Rabi}
\Om_{P,k}(t) = \a_k \OmP(t), \qquad \Om_{S,k}(t) = \b_k \OmS(t).
\ee
The dimensionless numbers $\a_k$ and $\b_k$ characterize
the relative strengths of the couplings while $\OmP(t)$ and $\OmS(t)$
are suitably chosen ``units'' of pump and Stokes Rabi frequencies.
We fix these ``units'' by choosing $\a_1=\b_1=1$, which means that
$\OmP(t)$ and $\OmS(t)$ are the pump and Stokes Rabi frequencies for state
$\state{1}$, $\OmP(t)\equiv\Om_{P,1}(t)$ and $\OmS(t)\equiv\Om_{S,1}(t)$,
and the constants $\a_k$ and $\b_k$ are the relative
strengths of the pump and Stokes couplings of state $\state{k}$
with respect to those for state $\state{1}$.
In real physical systems these constants contain
Clebsch-Gordan coefficients and Franck-Condon factors.
Moreover, without loss of generality we assume that $\OmP(t)$,
$\OmS(t)$, $\a_k$, and $\b_k$ are all positive.

\subsubsection{Adiabatic-transfer state}

As has been emphasized in \cite{Vitanov98multi,Vitanov98tune},
a necessary condition for complete adiabatic population transfer
in multistate systems is the existence of an adiabatic-transfer (AT)
state $\AT(t)$ which is defined as a nondegenerate
eigenstate\footnote{If the eigenstate $\AT(t)$ is degenerate,
there will be resonant population transfer to the other such
eigenstate(s) with a transition probability $\sin^2 \frac12 A$,
where $A$ is the pulse area of the nonadiabatic coupling between
these eigenstates, with inevitable population loss.
$A$ is independent on the pulse width $T$ and hence,
it does not vanish in the adiabatic limit $T\rightarrow\infty$,
unless the nonadiabatic coupling is identically equal to zero.}
of $\H(t)$ having the property
\be{AT}
\AT(t) = \left\{ \ba{ll} \state{i}, & t\rightarrow -\infty \\
			 \state{f}, & t\rightarrow +\infty \ea\right. ,
\ee
up to insignificant phase factors.
The dark state (\ref{dark}), which is a coherent superposition of states
$\state{i}$ and $\state{f}$ only, is the simplest example of such a state.
Under certain (quite restrictive) conditions it is an eigenstate
of the multi-$\Lambda$ system too, but in the general case
the Hamiltonian (\ref{H}) does not have such an eigenstate.
Under some more relaxed conditions, however, $\H(t)$ has an eigenstate
with the more general properties (\ref{AT}) (which allow for nonzero
transient contributions from the intermediate states),
and we derive these conditions below.


\section{The off-resonance case}

\label{Sec-off}

If all single-photon detunings are nonzero,
$\Delta_k\neq 0\ (k=1,2,\ldots,N)$, we have
\bml{detH}\bea
&& \det\H = \OmP^2 \OmS^2 \ \P
	\left( \Sa \Sb - \Sab^2 \right)\\
&&\qquad = \OmP^2 \OmS^2
	\sum_{k=1}^N \sum_{l=k+1}^N \P_{kl} (\a_k\b_l-\a_l\b_k)^2,
\emla
where
\be{sums}
 \Sa  = \!\sum_{k=1}^N \frac{\a_k^2}{\Delta_k}, \ \ 
 \Sb  = \!\sum_{k=1}^N \frac{\b_k^2}{\Delta_k}, \ \ 
 \Sab = \!\sum_{k=1}^N \frac{\a_k\b_k}{\Delta_k},\\
\ee
\be{products}
 \P = \prod_{k=1}^N \Delta_k,\ \ 
 \P_n = \prod_{k=1 \atop k \neq n}^N \Delta_k,\ \ 
 \P_{mn} = \prod_{k=1 \atop k \neq m,n}^N \Delta_k.
\ee
Hence, $\det\H \neq 0$ in the general case.
This means that, unlike in STIRAP ($N=1$),
$\H(t)$ does not necessarily have a zero eigenvalue.
We shall consider first in Sec.~\ref{Sec-zeroEV} the case when a zero
eigenvalue exists (with the anticipation, in analogy with STIRAP,
that the corresponding eigenstate is the desired AT state) and then
in Sec.~\ref{Sec-nonzeroEV} the case when it does not exist.


\subsection{A zero eigenvalue}

\label{Sec-zeroEV}

\subsubsection{Condition for a zero eigenvalue}

The condition for a zero eigenvalue is given by
\be{zeroEV}
\Sa \Sb - \Sab^2 
	= \sum_{k=1}^N \sum_{l=k+1}^N
	\frac{(\a_k\b_l-\a_l\b_k)^2}{\Delta_k\Delta_l} = 0.
\ee
Obviously, this condition depends only on the relative coupling
strengths and the detunings,
but neither on time nor on laser intensities.
Hence, it remains unchanged as the adiabatic limit is approached.

The eigenstate corresponding to the zero eigenvalue
most generally reads
\be{AT0}
\State{0}(t) = a_i(t)\state{i} + a_f(t)\state{f}
	+ \sum_{k=1}^N a_k(t)\state{k}.
\ee
The amplitudes of the intermediate states are given by
\be{a_k}
a_k(t) = -\frac{\Om_{P,k}(t)}{\Delta_k} a_i(t)
	 -\frac{\Om_{S,k}(t)}{\Delta_k} a_f(t),
\ee
with $k=1,2,\ldots,N$.
Obviously, they may be nonzero at finite times
but vanish at $\pm\infty$, $a_k(\pm\infty)=0$,
because both $\Om_{P,k}(t)$ and $\Om_{S,k}(t)$ vanish at infinity.
The amplitudes of the initial and final states satisfy both equations
\bml{ai-af}\bea
\label{EVEq1}
&& \Sa\OmP(t)a_i(t) + \Sab\OmS(t)a_f(t) = 0,\\
\label{EVEq2}
&& \Sab\OmP(t)a_i(t) + \Sb\OmS(t)a_f(t) = 0,
\emla
which are linearly dependent because of Eq.~(\ref{zeroEV}).

We are going to consider two cases:
when each term in the sum (\ref{zeroEV}) is zero
and when the individual terms may be nonzero but the total sum vanishes.


\subsubsection{Proportional couplings}

\label{Sec-proportional}

When
\be{proportional}
\frac{\a_1}{\b_1} = \frac{\a_2}{\b_2} = \ldots = \frac{\a_N}{\b_N}
 = 1,
\ee
each term in Eq.~(\ref{zeroEV}) vanishes and the zero eigenvalue
exists regardless of the detunings.
Condition (\ref{proportional}), which is essentially a condition on the
transition dipole moments, means that for each intermediate state
$\state{k}$, the ratio $\Om_{P,k}(t)/\Om_{S,k}(t)$ between the couplings to
states $\state{i}$ and $\state{f}$ is the same and does not depend on $k$. 
It follows from Eq.~(\ref{proportional}) that $\Sa=\Sb=\Sab\equiv \S$ and
we find from Eq.~(\ref{ai-af}) that $\OmP(t)a_i(t) + \OmS(t)a_f(t) = 0$,
provided $\S\neq 0$.
Then it follows from Eq.~(\ref{a_k}) that all
intermediate-state amplitudes are zero, $a_k(t)=0$.
It is easily seen that the zero-eigenvalue eigenstate coincides with
the dark state (\ref{dark}) and in the adiabatic limit it transfers
the population from state $\state{i}$ to state $\state{f}$,
bypassing the intermediate states.
Hence, when condition (\ref{proportional}) is fulfilled
the multi-$\Lambda$ system behaves very much like
the single $\Lambda$ system in STIRAP.
This confirms and generalizes the conclusions
in \cite{Coulston92,Carroll92} that complete population transfer
is possible when all $\a_k$ and $\b_k$ are equal.


\subsubsection{Arbitrary couplings}

\label{Sec-arbitrary}

Suppose now that condition (\ref{proportional}) is not fulfilled
while condition (\ref{zeroEV}) still holds.
This can be achieved by changing the two laser frequencies simultaneously
while maintaining the two-photon resonance,
which corresponds to adding a common detuning $\Delta$ to all single-photon
detunings, $\Delta_k \rightarrow \Delta_k+\Delta\ (k=1,2,\ldots,N)$.
For $N$ intermediate states, there are $N-2$ values of $\Delta$ for which
condition (\ref{zeroEV}) is satisfied.
In the zero-eigenvalue eigenstate (\ref{AT0})
the amplitudes of the intermediate states (\ref{a_k}) are generally nonzero.
However, if $\Sa$, $\Sb$, and $\Sab$ are nonzero
we have $|a_i(-\infty)|=1$ and $|a_f(+\infty)|=1$, and hence,
this eigenstate is an AT state, as defined by Eq.~(\ref{AT}).


\subsubsection{The case of vanishing $\Sa$, $\Sb$, and $\Sab$}

\label{Sec-000}

Let us now suppose that some of the sums $\Sa$, $\Sb$ and $\Sab$
are equal to zero.
If $\Sa \neq 0$ and $\Sb = 0$ (which also requires that $\Sab = 0$)
then it follows from Eq.~(\ref{EVEq1}) that $a_i(t) \equiv 0$.
Hence, we have $|a_f(\pm\infty)|=1$ and
$\State{0}(\pm\infty)=\state{f}$ (up to an irrelevant phase factor)
and there is no AT state.
A similar conclusion holds when $\Sa = 0$ and $\Sb \neq 0$: then
$a_f(t) \equiv 0$, $|a_i(\pm\infty)|=1$,
and $\State{0}(\pm\infty)=\state{i}$.

The case when $\Sa=\Sb=\Sab=0$ is different because then,
as can easily be shown, the Hamiltonian has two zero eigenvalues.
Consequently, there will be resonant transitions between the
corresponding degenerate eigenstates, even in the adiabatic limit,
which again prohibit complete STIRAP-like population transfer.

We shall illustrate this conclusion by calculating the final-state
population for proportional couplings (\ref{proportional}).
Due to the degeneracy, we have some freedom in choosing the
corresponding pair of orthonormal eigenstates.
As one of them, it is convenient to take state (\ref{dark}),
$\State{0}^{(1)}(t)=\dark(t)$,
because $\dark(-\infty)=\state{i}$ and $\dark(+\infty)=-\state{f}$.
The other zero-eigenvalue adiabatic state
can be determined by Gram-Schmidt ortogonalization and is
\be{AT1}
\State{0}^{(2)}(t) = \nu(t) \!\left[
	 \frac{\OmP(t)}{\Om(t)}\state{i}
	+\frac{\OmS(t)}{\Om(t)}\state{f}
	-\Om(t) \sum_{k=1}^N \frac{\a_k}{\Delta_k}\state{k} \right]\!\!,
\ee
with $\nu(t) = \left[ 1 + \Om^2(t) \sum_{k=1}^N \a_k^2 / \Delta_k^2
 \right] ^{-1/2}$,
where $\Om(t)$ is given by Eq.~(\ref{Omega}).
In the adiabatic limit, the population of state $\state{f}$,
which is equal to the probability of remaining in the adiabatic state
$\State{0}^{(1)}$, is given by
\be{Pf000}
P_f \approx \cos^2 \left[\int_{-\infty}^{\infty}
	\dot\vartheta(t) \nu(t) dt \right],
\ee
where $\dot\vartheta(t)=\dot\State{0}^{(1)}(t) \cdot \State{0}^{(2)}(t)$
and $\tan\vartheta(t)=\OmP(t)/\OmS(t)$.

In conclusion, when 
$\H(t)$ has a zero eigenvalue STIRAP-like transfer is possible only if
\be{zeroEV-ATS}
\Sa \neq 0,\qquad \Sb \neq 0, \qquad \Sab \neq 0.
\ee


\subsubsection{Examples}


\begin{figure}
\centerline{\psfig{width=75mm,file=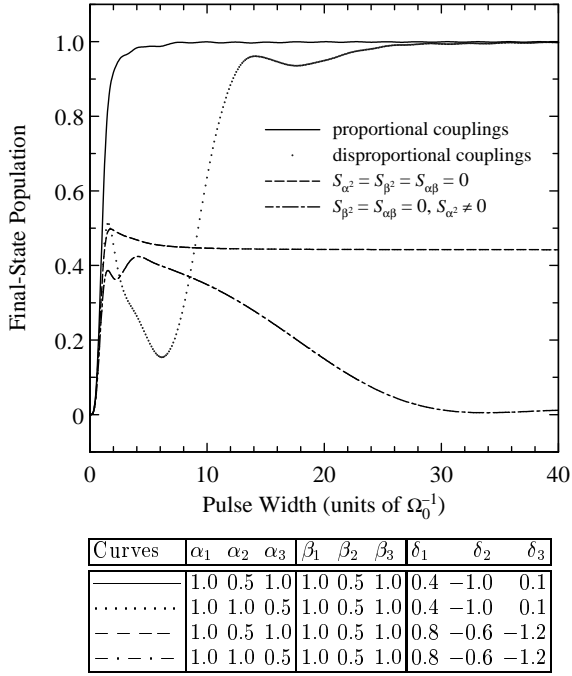}}
\vspace*{3mm}
\caption{
The final-state population $P_f$ as a function of the pulse width $T$
in the case of $N=3$ intermediate states for four combinations
of coupling strengths $\a_k$ and $\b_k$ and dimensionless detunings
$\d_k=\Delta_k/\Om_0$ given in the table.
}
\label{Fig-zeroEV}
\end{figure}


In Fig.~\ref{Fig-zeroEV}, the final-state population $P_f$
in the case of $N=3$ intermediate states is plotted against
the pulse width $T$ for four combinations of coupling strengths
and detunings.
The solid curve is for a case when condition (\ref{proportional})
is satisfied and the zero-eigenvalue eigenstate is the dark state
(\ref{dark}).
The dotted curve is for a case when condition (\ref{proportional})
is not satisfied but conditions (\ref{zeroEV}) and
(\ref{zeroEV-ATS}) are and the zero-eigenvalue eigenstate
is an AT state, Eq.~(\ref{AT}).
In both cases, the final-state population $P_f$ approaches unity as the
pulse width increases and the excitation becomes increasingly adiabatic.
The dashed curve is for a case when $\Sa=\Sb=\Sab=0$;
then, as the adiabatic limit is approached,
$P_f$ tends to the constant value $P_f\approx 0.442$,
predicted by Eq.~(\ref{Pf000}).
Finally, the dashed-dotted curve is for a case when $\Sb=\Sab=0$ and
$\Sa\neq 0$; then, in agreement with our analysis,
$P_f$ tends to zero in the adiabatic limit.


\begin{figure}
\centerline{\psfig{width=75mm,file=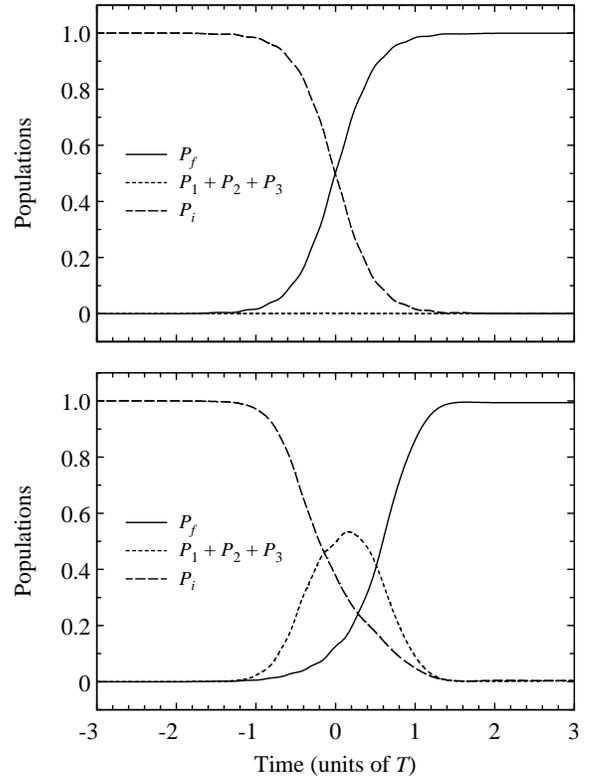}}
\vspace*{3mm}
\caption{
The time evolutions of the populations
in the case of $N=3$ intermediate states
for a pulse width $T=30\Om_0^{-1}$.
The interaction parameters for the upper and lower plots
are given respectively in the first and second rows
of the table in Fig.~\protect\ref{Fig-zeroEV}.
}
\label{Fig-time}
\end{figure}


In Fig.~\ref{Fig-time}, we show the time evolutions of the populations
in the case of $N=3$ intermediate states under almost adiabatic conditions.
The upper plot is for the solid-curve case in Fig.~\ref{Fig-zeroEV}
when 
the zero-eigenvalue eigenstate is the dark state (\ref{dark});
consequently, the intermediate states remain virtually unpopulated.
The lower plot is for the dotted-curve case in Fig.~\ref{Fig-zeroEV}
when the zero-eigenvalue eigenstate is an AT state with nonzero
components from the intermediate states; consequently, these states
acquire some transient populations.


\subsection{No zero eigenvalue}

\label{Sec-nonzeroEV}

When condition (\ref{zeroEV}) is not satisfied, $\det \H \neq 0$
and the Hamiltonian (\ref{H}) does not have a zero eigenvalue. 
This fact alone does not mean much because any chosen eigenvalue
can be made zero by shifting the zero energy level
with an appropriate time-dependent phase transformation.
More importantly, an AT state $\AT$, as defined by Eq.~(\ref{AT}),
may or may not exist and the conditions for its existence
are derived below.
The derivation is similar in spirit to that for multistate chains in
\cite{Vitanov98multi,Vitanov98tune}.

By setting $\OmP=\OmS=0$ in Eq.~(\ref{H}), we find that there are
{\it two} eigenvalues which vanish as $t\rightarrow \pm \infty$
(although they are nonzero at finite times),
while the others tend to the (nonzero) detunings $\Delta_k$.
At $\pm \infty$, each of the two eigenstates corresponding to the vanishing
eigenvalues is equal to either state $\state{i}$ or state $\state{f}$ or
a superposition of them.
Obviously, if an AT state exists, its eigenvalue $\EV_T$ should be one
of these eigenvalues.
Hence, we have to find the asymptotic behaviors of these eigenvalues
and the corresponding eigenstates, i.e., we need to determine {\it how}
the degeneracy of these eigenvalues is lifted by the laser fields.
We note here that only {\it one} eigenvalue vanishes when
$\OmP\rightarrow 0$ and $\OmS\neq 0$,
which happens at certain early times,
or when $\OmS\rightarrow 0$ and $\OmP\neq 0$,
which happens at certain late times.


\subsubsection{Early-time eigenvalues}

Let us first consider the case of early times ($t\rightarrow -\infty $).
It follows from the above remarks that as soon as the pulse
$\OmS$ arrives, one of the vanishing eigenvalues,
$\EV_l^-$ (the ``large'' one), departs from zero, while the other,
$\EV_s^-$ (the ``small'' one), remains zero until the pulse $\OmP$
arrives later.
Since $\EV_s^-$ vanishes when $\OmP\rightarrow 0$ and $\EV_l^-$
vanishes when $\OmS\rightarrow 0$, $\EV_s^-$ should be proportional
to some power of $\OmP$ and $\EV_l^-$ to some power of $\OmS$.
Since at those times $\OmP / \OmS \rightarrow 0$, the relation
$\left|\EV_s^-\right| \ll \left|\EV_l^-\right|$ holds;
hence, the names ``small'' and ``large''.

To determine $\EV_s^-$, we consider the eigenvalue equation,
\be{EVEq}
\det(\H-\EV{\bf 1}) =
	\h_0 + \h_1\EV + \ldots + \h_{N+2}\EV^{N+2} = 0,
\ee
as an implicit definition of the functional
dependence of $\EV_s^-$ on $\OmP$.
Note that $\h_0 \equiv \det\H$.
Since all $\h_k$ depend on $\OmP$ only via $\OmP^2$,
$\EV_s^-$ has a Taylor expansion in terms of $\OmP^2$.
We differentiate Eq.~(\ref{EVEq}) with respect to $\OmP^2$,
set $\OmP^2=0$ and $\EV_s^-(\OmP^2=0)=0$, and obtain
\be{derivative}
\h_0^\prime(0) + \h_1(0) \EV_s^{- \prime}(0) = 0, 
\ee
where a prime denotes $d/d\OmP^2$ and
\bml{dh0}\bea
&& \h_0^\prime(0) = \OmS^2 \P (\Sa\Sb - \Sab^2), \\
\label{h1}
&& \h_1(0) = \OmS^2 \P \Sb.
\emla
From here we find $\EV_s^{- \prime}(0)$,
replace it in the Taylor expansion of $\EV_s^-\left( \OmP^2\right)$,
and keeping the lowest-order nonzero term only, we obtain 
\be{Lms}
\EV_s^-\approx -\frac{\Sa\Sb-\Sab^2}{\Sb}\OmP^2.
\ee

In order to find the dependence of $\EV_l^-$ on $\OmS^2$, we set 
$\OmP=0$ in Eq.~(\ref{EVEq}), divide by $\EV$ (which amounts to 
removing the root $\EV_s^-$), differentiate with respect to $\OmS^2$,
set $\OmS^2=0$ and $\EV_l^-(\OmS^2=0)=0$, and find 
\be{Lml}
\EV_l^-\approx -\Sb\OmS^2.
\ee


\subsubsection{Late-time eigenvalues}

In a similar way, we find that at late times,
when $\OmS / \OmP \rightarrow 0$,
the two vanishing eigenvalues behave as 
\be{Lps}
\EV_s^+\approx -\frac{\Sa\Sb-\Sab^2}{\Sa}\OmS^2,
\ee
\be{Lpl}
\EV_l^+\approx -\Sa\OmP^2.
\ee


\subsubsection{Connectivity and AT condition}

It is easy to verify that the eigenstates corresponding to $\EV_s^-$ 
and $\EV_l^+$ coincide with state $\state{i}$, while those corresponding 
to $\EV_l^-$ and $\EV_s^+$ coincide with state $\state{f}$. 
Hence, the AT state $\AT$, if it exists, must have an eigenvalue that
coincides with $\EV_s^-$ at early times and with $\EV_s^+$ at late times.
It should be emphasized that $\EV_s^-$ and $\EV_s^+$ do {\it not}
necessarily correspond to the same eigenvalue and it may happen that
$\EV_s^-$ is linked to $\EV_l^+$ rather than $\EV_s^+$;
then an AT state does not exist.
In any case, the upper (the lower) of the two eigenvalues at $-\infty$ is
connected to the upper (the lower) of the two eigenvalues at $+\infty $.
Since $\left| \EV_l^-\right| \gg \left| \EV_s^-\right| $
 and $\left| \EV_l^+\right| \gg \left|\EV_s^+\right| $,
the linkage is determined by the signs of the
``large'' eigenvalues $\EV_l^-$ and $\EV_l^+$. 
If they have the same signs, they will be both above (or below) 
$\EV_s^-$ and $\EV_s^+$ and hence, the desired linkages 
$\EV_l^-\leftrightarrow \EV_l^+$ and 
$\EV_s^-\leftrightarrow \EV_s^+$ will take place. 
If $\EV_l^- $ and $\EV_l^+$ have opposite signs they cannot 
be connected because such an eigenvalue will cross the one linking 
$\EV_s^-$ and $\EV_s^+$, which is impossible. 
Thus, from this analysis and Eqs.~(\ref{Lml}) and (\ref{Lpl}) we conclude 
that the necessary and sufficient {\it condition for existence of 
an adiabatic-transfer state} is 
\be{ATcondition}
\Sa\Sb > 0. 
\ee

It is easy to see that condition (\ref{zeroEV}) for existence of a
zero eigenvalue, along with condition (\ref{zeroEV-ATS}) for existence
of AT state in this case, agree with condition (\ref{ATcondition}).
Indeed, we have $\Sa\Sb = \Sab^2 > 0$.
Moreover, the zero eigenvalue is reproduced correctly by
Eqs.~(\ref{Lms}) and (\ref{Lps}).

\subsubsection{The case of vanishing $\Sa$ and $\Sb$}

The derivation of the AT condition (\ref{ATcondition}) suggests that
both sums $\Sa$ and $\Sb$ should be nonzero.
Let us examine the case when one of them is zero, e.g., $\Sb=0$.
By going through the derivation that leads to Eq.~(\ref{Lms})
we find that now, as evident from Eq.~(\ref{h1}), we have $\h_1(0)=0$.
It follows from Eq.~(\ref{EVEq}) that now two, rather than one,
eigenvalues vanish when $\OmP = 0$ and $\OmS\neq 0$
at early times (because then $\h_0 \equiv \det\H = 0$ too).
Hence, in contrast to the case of $\Sb\neq 0$,
the arrival of the Stokes pulse $\OmS(t)$ does not make one of these
eigenvalues depart from zero but rather their degeneracy is lifted only
with the arrival of the pump pulse $\OmP(t)$ later.
The implication is that the initial state $\state{i}$ cannot be
identified with a single adiabatic state at $t\rightarrow -\infty$
but it is rather equal to a superposition of two adiabatic states%
\footnote{The initial state $\state{i}$ is associated with
a single adiabatic state at $-\infty$
when the arrival of the pump pulse lifts the degeneracy of
one and only one eigenvalue.
Similarly, the final state $\state{f}$ is associated with
a single adiabatic state at $+\infty$
when the vanishing Stokes pulse restores the degeneracy of
one and only one eigenvalue.
}; hence, there is no AT state.
A similar conclusion holds in the case when $\Sa=0$:
then the final state $\state{f}$ cannot be identified with a single
adiabatic state at $t\rightarrow +\infty$.
Therefore, for $\Sa=0$ or $\Sb=0$ an AT state does not exist, as follows
formally from Eq.~(\ref{ATcondition}).

The situation with the mixed sum $\Sab$ is different.
In the above derivation
the condition $\Sab \neq 0$ was not required
anywhere which means that an AT state exists even for $\Sab = 0$.
We will return to this problem in Sec.~\ref{Sec-AE}.


\subsubsection{Examples}


\begin{figure}[tb]
\centerline{\psfig{width=84mm,file=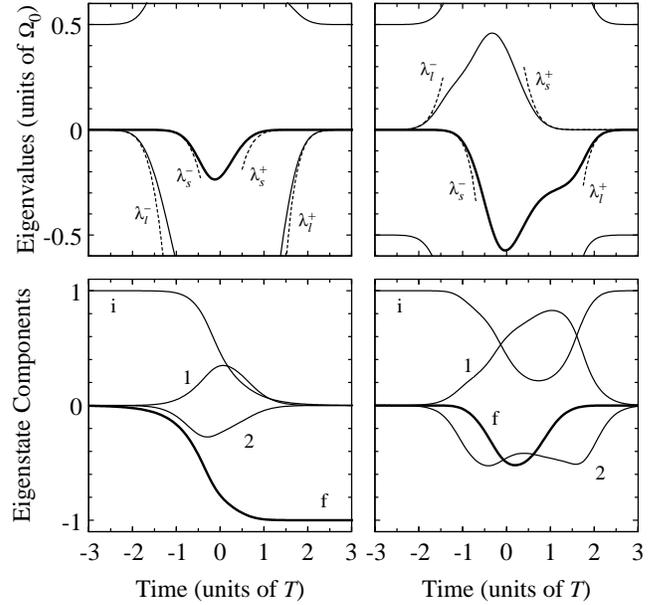}}
\vspace*{3mm}
\caption{
The upper row of figures show the time evolutions of the eigenvalues in
the case of $N=2$ intermediate states for two sets of detunings: 
$\Delta_1 = 0.5\Om_0, \Delta_2 = 1.5\Om_0$ (left figure) and
$\Delta_1 = -0.5\Om_0, \Delta_2 = 0.5\Om_0$ (right figure).
The coupling strengths are the same in both cases,
$\a_1=\b_1=1$, $\a_2=2$, and $\b_2=0.5$.
The solid curves are calculated numerically and the dashed curves show
the asymptotic approximations (\protect\ref{Lms})-(\protect\ref{Lpl}).
The bottom row of figures show the components of this eigenstate which
is equal to the bare state $\state{i}$ initially.
It corresponds to the eigenvalue which is shown by a thick curve
in the corresponding upper figure and whose asymptotics
at early times is described by $\EV_s^-$.
The labels denote the bare states to which the components belong and  
the final-state component is shown by a thick curve.
}
\label{Fig-EV}
\end{figure}


In Fig.~\ref{Fig-EV} we have plotted the time evolutions of the eigenvalues
(upper row of figures) in the case of $N=2$ intermediate states for
two combinations of detunings and the same set of coupling strengths.
The solid curves are calculated numerically and the dashed curves show
our asymptotic approximations (\protect\ref{Lms})-(\protect\ref{Lpl}).
The bottom row of figures show the components of this eigenstate which
is equal to bare state $\state{i}$ initially;
it corresponds to the eigenvalue whose asymptotics
at early times is described by $\EV_s^-$.
Hence, the squared components of this eigenstate give the populations
of the four bare states in the adiabatic limit.
As we can see in the {\it left column} of figures,
there is an eigenvalue whose asymptotics is given
by $\EV_s^-$ at early times and by $\EV_s^+$ at late times; this is
so because condition (\ref{ATcondition}) is satisfied in this case.
The corresponding eigenstate is an AT state,
as evident from the bottom left figure, because it is equal
to state $\state{i}$ initially and to state $-\state{f}$ finally.
For the case shown in the {\it right column} of figures,
there is no AT eigenvalue because the asymptotic behaviors
$\EV_s^-$ and $\EV_s^+$ are related to two different eigenvalues;
this is so because condition (\ref{ATcondition})
is not satisfied in this case.
Consequently, there is no AT eigenstate, as evident from the bottom
right figure, because the shown eigenstate is equal
to state $\state{i}$ both initially and finally.


\begin{figure}[tbp]
\centerline{\psfig{width=75mm,file=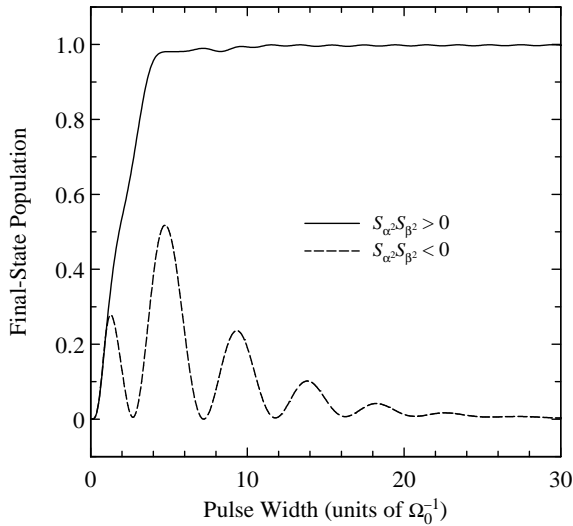}}
\vspace*{3mm}
\caption{
The final-state population $P_f$ as a function of the pulse width $T$ in
the case of $N=2$ intermediate states for the same two sets of
interaction parameters as in Fig.~\protect\ref{Fig-EV}.
The solid and dashed curves correspond to the left and right columns
in Fig.~\protect\ref{Fig-EV}, respectively.
The detunings for the two curves are
$\Delta_1 = 0.5\Om_0$, $\Delta_2 = 1.5\Om_0$ (solid curve), and
$\Delta_1 = -0.5\Om_0$, $\Delta_2 = 0.5\Om_0$ (dashed curve).
The coupling strengths are the same for both curves,
$\a_1=\b_1=1$, $\a_2=2$, and $\b_2=0.5$.
}
\label{Fig-2S-T}
\end{figure}


In Fig.~\ref{Fig-2S-T} we have plotted the final-state population
$P_f$ as a function of the pulse width $T$ in the case of $N=2$
intermediate states for the same two sets of interaction parameters
as in Fig.~\ref{Fig-EV}.
The solid and dashed curves in Fig.~\ref{Fig-2S-T} correspond
to the left and right columns in Fig.~\ref{Fig-EV}, respectively.
As follows from Eq.~(\ref{ATcondition}), an AT state exists for
the solid curve and does not exist for the dashed curve.
Indeed, as seen in the figure, as $T$ increases,
the final-state population $P_f$ approaches unity for the solid
curve and zero for the dashed curve.


\begin{figure}[tbp]
\centerline{\psfig{width=85mm,file=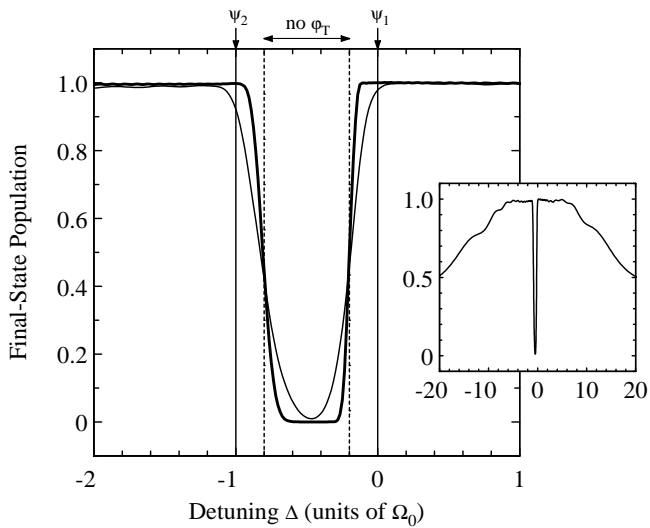}}
\vspace*{3mm}
\caption{
The final-state population $P_f$ as a function of
the single-photon detuning $\Delta$
of the pump and Stokes fields from the lowest intermediate state.
We have taken two intermediate states $\state{1}$ and $\state{2}$
with $\Delta_1=\Delta$, $\Delta_2=\Om_0 + \Delta$.
The positions of the single-photon resonances with these states
are shown by the vertical solid lines.
The region where an AT state does not exist,
$-0.8\Om_0 \protect{\leq} \Delta \protect{\leq} -0.2\Om_0$,
is shown by vertical dashed lines.
The relative coupling strengths are taken as in
Figs.~\protect\ref{Fig-EV} and \protect\ref{Fig-2S-T},
$\a_1=\b_1=1$, $\a_2=2$, and $\b_2=0.5$.
The thin curve is for $\Om_0 T = 20$
and the thick curve is for $\Om_0 T = 80$.
The inset shows $P_f$ for $\Om_0 T = 20$ in a wider range.
}
\label{Fig-2S-D}
\end{figure}


In Fig.~\ref{Fig-2S-D} the final-state population $P_f$ is plotted
as a function of the single-photon detuning $\Delta$
of the pump and Stokes fields from the lowest intermediate state.
We have taken two intermediate states $\state{1}$ and $\state{2}$
with $\Delta_1=\Delta$, $\Delta_2=\Om_0 + \Delta$.
The coupling strengths are taken the same as in
Figs.~\ref{Fig-EV} and \ref{Fig-2S-T}.
In the region where an AT state does not exist,
$-0.8\Om_0 \leq \Delta \leq -0.2\Om_0$
[calculated from Eq.~(\ref{ATcondition})],
the transfer efficiency is low,
while outside it the transfer efficiency is almost unity,
as a result of the existence of an AT state.
The left and right columns of plots in Fig.~\ref{Fig-EV}
(respectively, the solid and dashed curves in Fig.~\ref{Fig-2S-T})
correspond to detunings $\Delta=0.5\Om_0$ and $\Delta=-0.5\Om_0$,
respectively.
The inset shows how the transfer efficiency eventually decreases at
large detunings which, as in STIRAP \cite{Vitanov97D},
is due to deteriorating adiabaticity.


\begin{figure}[tb]
\centerline{\psfig{width=75mm,file=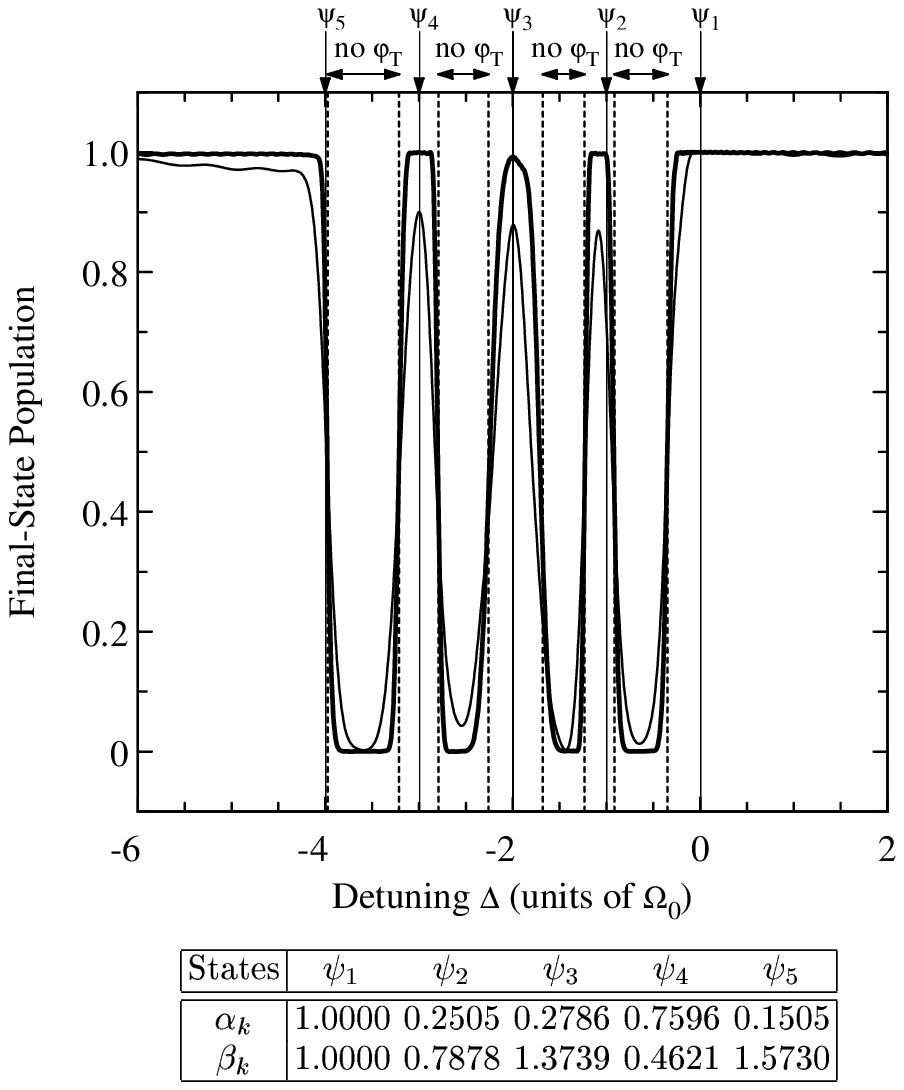}}
\vspace*{3mm}
\caption{
The final-state population $P_f$ as a function of
the single-photon detuning $\Delta$
of the pump and Stokes fields from the lowest intermediate state.
We have taken $N=5$ equidistant intermediate states, i.e.,
$\Delta_k=(k-1)\Om_0 + \Delta$ with $k=1,2,3,4,5$.
The positions of the single-photon resonances with these states
are shown by the vertical solid lines.
The relative coupling strength $\a_k$ and $\b_k$ are taken randomly,
as provided by the random generator of our computer, and given
in the table (up to four digits after the decimal point).
The thin curve is for $\Om_0 T = 20$
and the thick curve is for $\Om_0 T = 80$.
The vertical dashed lines separate the regions where an AT state exists
or does not exist, as calculated from Eq.~(\protect\ref{ATcondition}).
}
\label{Fig-5S-D}
\end{figure}


In Fig.~\ref{Fig-5S-D}, we have plotted the final-state population
$P_f$ as a function of the single-photon detuning $\Delta$
of the pump and Stokes fields from the lowest intermediate state
for the case of $N=5$ equidistant intermediate states and
randomly taken coupling strengths $\a_k$ and $\b_k$.
As Fig.~\ref{Fig-5S-D} shows, there are three distinct domains
of single-photon detunings: $\Delta \lesssim -4\Om_0$,
$-4\Om_0\lesssim \Delta \lesssim 0$, and $\Delta \gtrsim 0$.
For $\Delta \lesssim -4\Om_0$ and $\Delta \gtrsim 0$,
we have $P_f \approx 1$,
whereas for $-4\Om_0\lesssim \Delta \lesssim 0$, there are alternative
regions of high and low transfer efficiency.
This behavior is easily explained by the AT condition (\ref{ATcondition}).
As $\Delta$ changes, we pass through the zero points of the sums $\Sa(\Delta)$
and $\Sb(\Delta)$, thus going from an interval where these sums have the same
sign (where $P_f \approx 1$) to an interval where they have opposite signs
(where $P_f \approx 0$) and vice versa.
Obviously, for sufficiently large and negative $\Delta$,
both $\Sa(\Delta)$ and $\Sb(\Delta)$ are always negative
and condition (\ref{ATcondition}) is satisfied, which ensures the
existence of AT state and STIRAP-like unit transfer efficiency.
Similarly, for sufficiently large and positive $\Delta$,
both $\Sa(\Delta)$ and $\Sb(\Delta)$ are always positive
and we have unit transfer efficiency there too.
The conclusion is that whenever the pump and Stokes lasers are tuned
below or above all intermediate states, STIRAP-like transfer is always
guarantied in the adiabatic limit.
When the lasers are tuned within the manifold of intermediate states
such a transfer may or may not take place, depending on whether the AT
condition (\ref{ATcondition}) is satisfied or not.
Furthermore, tuning the lasers below or above all states appears
to be more reasonable even than tuning on resonance with
an intermediate state because adiabaticity is achieved more easily,
as the curves for $\Om_0 T = 20$ in Figs.~\ref{Fig-2S-D}
and \ref{Fig-5S-D} show.
Moreover, as follows from Eq.~(\ref{a_k}), the transient
intermediate-state populations $P_k(t)$ decrease
with the detuning as $\Delta_k^{-2}$.


\section{A resonant intermediate state}

\label{Sec-on}

\subsection{A zero eigenvalue}

If a certain detuning is equal to zero, $\Delta_n=0$, we have
\bml{detH(0)}\bea
\label{detH(0)-1}
&& \det\H = \OmP^2 \OmS^2 \sum_{k=1 \atop k\neq n} ^N
	\P_{nk} (\a_k\b_n-\a_n\b_k)^2\\
\label{detH(0)-2}
&&\qquad = \OmP^2 \OmS^2 \P_n\!
  \left[ \a_n^2\Sbn \!-\! 2\a_n\b_n\Sabn \!+\! \b_n^2\San \right]\!,
\emla
where the $\S^{(n)}$-sums are defined as the $\S$-sums but without the
$n$-th terms,
\be{sums-n}
 \San  = \!\sum_{k=1 \atop k\neq n}^N \frac{\a_k^2}{\Delta_k}, \ \ 
 \Sbn  = \!\sum_{k=1 \atop k\neq n}^N \frac{\b_k^2}{\Delta_k}, \ \ 
 \Sabn = \!\sum_{k=1 \atop k\neq n}^N \frac{\a_k\b_k}{\Delta_k},\\
\ee

The intermediate-state amplitudes, except $a_n(t)$,
are given by Eq.~(\ref{a_k}).
The equation for $a_n(t)$ is replaced by
\be{nth-eqn}
\Om_{P,n}(t) a_i(t) + \Om_{S,n}(t) a_f(t) = 0,
\ee
while Eqs.~(\ref{ai-af}) are replaced by
\bml{ai-af-0}
\bea
\label{EVEq1-0}
&& \San\OmP(t)a_i(t) + \Sabn\OmS(t)a_f(t) + \a_n a_n(t) = 0,\\
\label{EVEq2-0}
&& \Sabn\OmP(t)a_i(t) + \Sbn\OmS(t)a_f(t) + \b_n a_n(t) = 0.
\emla

We can again consider the two possibilities: when condition
(\ref{proportional}) is fulfilled and when it is not.
In the case of {\it proportional couplings} (\ref{proportional}),
it can readily be shown that the zero-eigenvalue adiabatic state
is given again by the dark state (\ref{dark}).
Moreover, now condition (\ref{zeroEV-ATS}) is not required and the dark
state $\dark(t)$ is a zero-eigenvalue eigenstate of $\H(t)$ even when
$\San$,  $\Sbn$, and $\Sabn$ vanish%
\footnote{Note that unlike the off-resonance case with $\Sa=\Sb=\Sab=0$
(Sec.~\ref{Sec-000}), no additional zero eigenvalues exist for
$\Delta_n=0$ when $\San=\Sbn=\Sabn=0$.}.
In the case of {\it arbitrary couplings}, the sum (\ref{detH(0)-1})
can vanish only by accident because we are not allowed
to "scan" the pump and Stokes laser frequencies across the intermediate
states as we would violate the assumed single-photon resonance condition
$\Delta_n=0$.
If this happens it can easily be shown that again, the zero-eigenvalue
eigenstate is an AT state, the only difference from the
off-resonance case (Sec.~\ref{Sec-arbitrary})
 being that now the AT state does not contain
a component from the resonant bare state, $a_n(t)=0$.
In both cases, we have complete population transfer in the adiabatic limit.


\subsection{Nonzero eigenvalue}

For $\det \H \neq 0$, we follow the same approach as for nonzero
detunings (Sec.~\ref{Sec-nonzeroEV}).
By setting $\OmP=\OmS=0$ in Eq.~(\ref{H}) we find that for $\Delta_n=0$,
there are {\it three}, rather than two, vanishing eigenvalues.
Hence, we have to establish how the new, third, zero eigenvalue
affects the AT state.


\subsubsection{Early-time eigenvalues}

It is readily seen from Eq.~(\ref{H}) that only one eigenvalue vanishes
for $\OmP=0$ and $\OmS\neq 0$.
This means that as soon as the Stokes pulse $\OmS(t)$ arrives, the
degeneracy of two of the eigenvalues, $\EV_{l,1}^-$ and $\EV_{l,2}^-$,
is lifted and they depart from zero, while the third eigenvalue
$\EV_s^-$ stays zero until the pulse $\OmP(t)$ arrives later.
Hence, there are two ``large'' eigenvalues
and one ``small'' eigenvalue.

The ``small'' eigenvalue $\EV_s^-(\OmP^2)$ can be determined in the same
manner as for nonzero detunings.
We have
\bml{ES2}\bea
&& \h_0^\prime(0) = \OmS^2 \P_n
	\left[\a_n^2\Sbn - 2\a_n\b_n\Sabn + \b_n^2\San \right], \\
&& \h_1(0) = \OmS^2 \b_n^2 \P_n.
\emla
Using Eq.~(\ref{derivative}), we find $\EV_s^{- \prime}(0)$
and obtain 
\be{Lms0}
\EV_s^- \approx -\frac 1{\b_n^2}
\left[\a_n^2\Sbn - 2\a_n\b_n\Sabn + \b_n^2\San \right] \OmP^2.
\ee

In order to determine the other two eigenvalues
$\EV_{l,1}^-$ and $\EV_{l,2}^-$, which depart from zero with $\OmS$,
we set $\OmP=0$ in Eq.~(\ref{EVEq}) and divide by $\EV$
(thus removing the root $\EV_s^-$).
Keeping the terms of lowest order with respect to $\OmS$ and $\EV$,
we find that $-\OmS^2 \b_n^2 + \EV^2 \approx 0$, and we identify
$\EV_{l,1}^-$ and $\EV_{l,2}^-$ as the two roots of this equation,
\be{Lml0}
\EV_{l,1}^- \approx - \b_n \OmS, \qquad
\EV_{l,2}^- \approx \b_n \OmS.
\ee


\subsubsection{Late-time eigenvalues}

In a similar fashion, we find that for $t\rightarrow +\infty $, the three
vanishing eigenvalues behave as
\be{Lps0}
\EV_s^+ \approx -\frac 1{\a_n^2}
\left[\a_n^2\Sbn - 2\a_n\b_n\Sabn + \b_n^2\San \right] \OmS^2,
\ee
\be{Lpl0}
\EV_{l,1}^+ \approx - \a_n \OmP, \qquad
\EV_{l,2}^+ \approx \a_n \OmP.
\ee


\subsubsection{Connectivity}

It is straightforward to show that the eigenstate associated with
$\EV_s^-$ tends to state $\state{i}$ as $t\rightarrow -\infty$
and the eigenstate associated with $\EV_s^+$
tends to state $\state{f}$ as $t\rightarrow +\infty$.
The eigenstates corresponding to the ``large'' eigenvalues tend to
superpositions of states $\state{f}$ and $\state{n}$ initially
and to superpositions of states $\state{i}$ and $\state{n}$ finally.
Hence, if $\EV_s^-$ and $\EV_s^+$ correspond to the same eigenvalue,
the corresponding eigenstate will be the desired AT state $\AT(t)$.
We have seen above that in the general off-resonance case,
this may or may not take place.
In the present case of a single-photon resonance, however,
this is always the case.
To show this we first note that the eigenvalues,
which do not vanish at $\pm \infty $,
do not interfere in the linkages between the vanishing eigenvalues
because each of the nonvanishing eigenvalues $\EV_k(t)$ tends to
the corresponding detuning $\Delta_k$ at both $\pm \infty $.
Hence, the eigenvalues that are above (below) the three vanishing
eigenvalues at $-\infty$ remain above (below) them at $+\infty$ as well.

Let us now consider the linkages between the three vanishing eigenvalues.
Insofar as $\OmP/\OmS\rightarrow 0$ as $t\rightarrow -\infty $, we have
$\EV_{l,1}^-<\EV_s^-<\EV_{l,2}^-$.
Also, since $\OmS /\OmP \rightarrow 0$ as $t \rightarrow +\infty$, we have
$\EV_{l,1}^+<\EV_s^+<\EV_{l,2}^+$.
This means that the linkages
$\EV_{l,1}^- \leftrightarrow \EV_{l,1}^+$,
$\EV_s^- \leftrightarrow \EV_s^+$, and
$\EV_{l,2}^- \leftrightarrow \EV_{l,2}^+$
take place, and therefore, the AT state $\AT(t)$ always exists
when the lasers are tuned to resonance with an intermediate state,
which is indeed seen in Figs.~\ref{Fig-2S-D} and \ref{Fig-5S-D}.


\subsection{Examples}


\begin{figure}[tb]
\centerline{\psfig{width=84mm,file=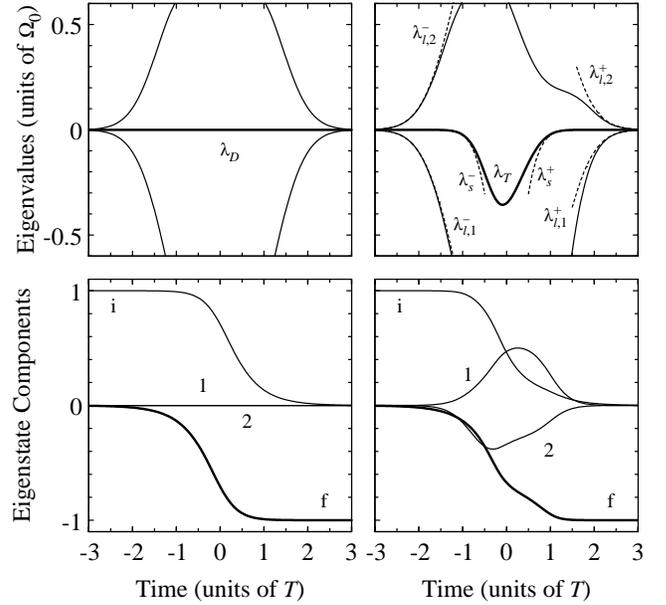}}
\vspace*{3mm}
\caption{
The upper row of figures show the time evolutions
of the three eigenvalues that vanish at $\pm\infty$
in the case of $N=2$ intermediate states
for two combinations of coupling strengths:
$\a_1=\b_1=1$, $\a_2=\b_2=0.5$ (upper left figure) and
$\a_1=\b_1=1$, $\a_2=2$, and $\b_2=0.5$ (upper right figure).
The detunings are the same in both cases:
$\Delta_1 = 0, \Delta_2 = \Om_0$.
The solid curves are calculated numerically
and the dashed curves show our asymptotic approximations
(\protect\ref{Lms0})-(\protect\ref{Lpl0}).
The bottom row of figures show the components of this eigenstate
which is equal to the bare state $\state{i}$ initially.
It corresponds to the eigenvalue which is shown by a thick curve
in the corresponding upper figure.
This is the zero eigenvalue $\lambda_D$ for the left figure
and the eigenvalue $\lambda_T$ whose asymptotics at early times is
described by $\EV_s^-$ for the right figure.
The labels denote the bare states to which the components belong and  
the final-state component is shown by a thick curve.
}
\label{Fig-EV0}
\end{figure}


In Fig.~\ref{Fig-EV0} we have plotted the time evolutions
of the three eigenvalues that vanish at $\pm\infty$
(upper row of figures) in the case of $N=2$ intermediate states for
two sets of coupling strengths and the same detunings,
$\Delta_1 = 0, \Delta_2 = \Om_0$, i.e., the lower intermediate
state is on single-photon resonance.
The top left figure is for proportional coupling strengths
(\ref{proportional}) which give rise to a zero eigenvalue $\EV_D$
and correspondingly, to a dark state $\dark(t)$.
The top right figure is for a case when Eq.~(\ref{proportional})
is not satisfied and there is no zero eigenvalue.
Note the pair of eigenvalues $\EV_{l,1}^-$ and $\EV_{l,2}^-$
which depart from zero in opposite directions
with the arrival of the Stokes pulse at early times
and the pair of eigenvalues $\EV_{l,1}^+$ and $\EV_{l,2}^+$
which vanish with the disappearence of the pump pulse at late times.
The bottom row of figures show the components of this eigenstate which
is equal to the bare state $\state{i}$ initially;
it corresponds to the zero eigenvalue for the left figure
and to the eigenvalue whose asymptotics at early times
is described by $\EV_s^-$ for the right figure.
The squared components of this eigenstate give the populations
of the four bare states in the adiabatic limit.
As predicted by our analysis, the AT state is seen to exist in both cases.
The difference is that for the left column of figures,
the AT state is the dark state (\ref{dark}), which does not contain
components from the intermediate states, while for the right column
of figures, the AT state contains such components.


\begin{figure}[tbp]
\centerline{\psfig{width=75mm,file=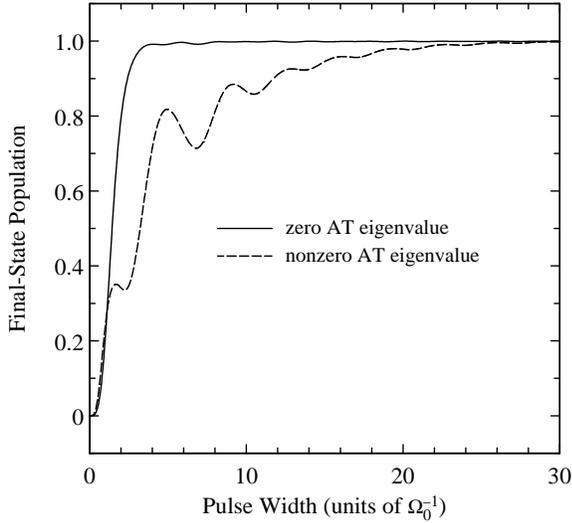}}
\vspace*{3mm}
\caption{
The final-state population $P_f$ as a function of the pulse width $T$
in the case of $N=2$ intermediate states for the same two sets of
interaction parameters as in Fig.~\protect\ref{Fig-EV0}.
The solid and dashed curves correspond to the left and right columns
in Fig.~\protect\ref{Fig-EV0}, respectively.
The detunings are the same in both cases:
$\Delta_1 = 0, \Delta_2 = \Om_0$.
The solid curve is for coupling strengths
$\a_1=\b_1=1$, $\a_2=\b_2=0.5$ (then the AT eigenvalue is $\EV_D=0$),
and the dashed curve is for $\a_1=\b_1=1$, $\a_2=2$, and $\b_2=0.5$
(then the AT eigenvalue is nonzero).
}
\label{Fig-2S-T0}
\end{figure}


In Fig.~\ref{Fig-2S-T0}, we have plotted the final-state population
$P_f$ as a function of the pulse width $T$ in the case of $N=2$
intermediate states for the same two sets of interaction parameters
as in Fig.~\ref{Fig-EV0}.
The solid and dashed curves in Fig.~\ref{Fig-2S-T0} correspond
to the left and right columns in Fig.~\ref{Fig-EV0}, respectively.
In agreement with our conclusions, an AT state exists in both cases
and the final-state population $P_f$ approaches unity as $T$ increases.


\section{Degenerate resonant intermediate states}

\label{Sec-degenerate}

Let us suppose now that $N_0$ detunings vanish, i.e., that there are
$N_0$ degenerate resonant intermediate states,
and let us assume without loss of generality that these states are
$\state{1},\state{2},\ldots,\state{N_0}$.
If two of the detunings are equal to zero, $\Delta_1=\Delta_2=0$, we have
\be{detH00}
\det\H = \OmP^2 \OmS^2 \P_{12} (\a_1\b_2-\a_2\b_1)^2,
\ee
and hence, a zero eigenvalue exists when $\a_1/\b_1=\a_2/\b_2$.
If three or more detunings are equal to zero then $\det\H\equiv 0$,
and a zero eigenvalue always exists,
with no restrictions on the interaction parameters.


\subsection{Proportional couplings}

In the zero-eigenvalue eigenstates, the components $a_i(t)$ and
$a_f(t)$ of states $\state{i}$ and $\state{f}$ satisfy the equations
\be{ampl}
\a_n\OmP(t)a_i(t) + \b_n\OmS(t)a_f(t) = 0,
\ee
with $n=1,2,\ldots,N_0$.
A nonzero solution for $a_i(t)$ and $a_f(t)$ requires that
\be{proportional0}
\frac{\a_1}{\b_1} = \frac{\a_2}{\b_2} = \ldots
	 = \frac{\a_{N_0}}{\b_{N_0}} = 1.
\ee
Otherwise, a zero-eigenvalue eigenstate cannot be an AT state.
If relation (\ref{proportional0}) is fulfilled,
the amplitudes of the degenerate states are linearly dependent%
\footnote{
This is so because the function $f(t)=\a_nc_1(t)-\a_1c_n(t)$
satisfies the differential equation $df(t)/dt = 0$
with the initial condition $f(-\infty) = 0$; hence, $f(t)=0$.}
\be{lindep}
c_n(t) = c_1(t) \frac{\a_n}{\a_1},
\qquad (n=2,3,\ldots,N_0).
\ee
This relation allows to replace in the Schrödinger equation (\ref{SEq})
the probability amplitudes of the degenerate states
by an effective amplitude given by
\be{ceff}
c_{{\rm eff}}(t) = \mu c_1(t),
\ee
with pump and Stokes Rabi frequencies given by
\be{PS}
\Om_{P,{\rm eff}}(t) = \mu\OmP(t),\qquad
\Om_{S,{\rm eff}}(t) = \mu\OmS(t),
\ee
where $\mu = (1/\a_1)\sqrt{\a_1^2 + \a_2^2 + \ldots + \a_{N_0}^2}$.
Thus the original problem with $N_0$ resonant states is reduced to
an equivalent problem involving a single resonant state.
As we pointed out in Sec.~\ref{Sec-on}, an AT state $\AT(t)$
always exists in this case.
Moreover, in the case when all couplings (and not only those for
the degenerate states) are proportional
the AT state is the dark state $\dark(t)$.


\subsection{Arbitrary couplings}

If Eq.~(\ref{proportional0}) is not fulfilled, the zero-eigenvalue
eigenstate(s) cannot be an AT state because the components
$a_i(t)$ and $a_f(t)$ from states $\state{i}$ and $\state{f}$ vanish.
There still might be a possibility that one of the two nonzero
eigenvalues, which vanish at $\pm \infty$, corresponds to an AT state.
We shall show, however, that this is not the case.

For $N_0$ degenerate resonant states, it can readily be shown that
the number of zero eigenvalues is
$N_0-2$ for $\OmP(t)\neq 0$ and $\OmS(t)\neq 0$,
$N_0$ for $\OmP(t) = 0$ and $\OmS(t)\neq 0$
[or for $\OmP(t) \neq 0$ and $\OmS(t) = 0$],
and $N_0+2$ for $\OmP(t) = \OmS(t) =0$.
This means that when the Stokes pulse arrives at early times, it
lifts the degeneracy of two of the $N_0+2$ zero eigenvalues.
When the pump pulse arrives later, if lifts the degeneracy
of another pair of the remaining $N_0$ eigenvalues.
The reverse process occurs at large positive times.
The remaining $N_0-2$ zero eigenvalues stay degenerate all the time.
The implication is that the initial state $\state{i}$
and the final state $\state{f}$ are given by superpositions of
eigenstates both at $-\infty$ and $+\infty$,
which means that an AT does not exist.


\section{Adiabatic elimination of the off-resonance states}

\label{Sec-AE}

\subsection{The off-resonance case}

An insight into the population transfer process can be obtained from the
adiabatic-elimination approximation.
When the single-photon detuning $\Delta_k$ of a given intermediate state
$\state{k}$ is large compared to the couplings $\Om_{P,k}$ and
$\Om_{S,k}$ of this state to states $\state{i}$ and $\state{f}$,
this state can be eliminated adiabatically by setting $dc_k/dt=0$
and expressing $c_k$ from the resulting algebraic equation.
By adiabatically eliminating all intermediate states the general
$(N+2)$-state problem is reduced to an effective two-state problem
for the initial and final states,
\be{AE2}
i \frac{d}{dt} \bM{ c_i \cr c_f }\eM \approx
\bM{
\OmP^2\Sa    & \OmP\OmS\Sab \cr
\OmP\OmS\Sab & \OmS^2\Sb
}\eM
\bM{ c_i \cr c_f }\eM.
\ee

The ``detuning'' in this two-state problem is
$\Deff(t) = \OmS^2(t)\Sb - \OmP^2(t)\Sa$.
Obviously, if $\Sa\Sb > 0$, $\Deff(t)$ has different signs
at $\pm\infty$ and the transition is of level-crossing type,
while if $\Sa\Sb < 0$, $\Deff(t)$ has the same sign
at $\pm\infty$ and there is no crossing.
Hence, in the adiabatic limit, the transition probability from
state $\state{i}$ to state $\state{f}$ will be unity for $\Sa\Sb > 0$
and zero for $\Sa\Sb < 0$,
in agreement with the AT condition (\ref{ATcondition}).

The ``coupling'' in the effective two-state problem (\ref{AE2}) is
$\Omeff(t) = \OmP(t)\OmS(t)\Sab$.
Obviously, it vanishes for $\Sab=0$ which suggests that there is no
transition from state $\state{i}$ to state $\state{f}$,
both for $\Sa\Sb > 0$ and $\Sa\Sb < 0$.
However, we know from Sec.~\ref{Sec-000} that this prediction
is incorrect and that an AT exists even in this case,
as long as $\Sa\Sb > 0$.
This somewhat surprising discrepancy derives from the fact that
for $\Sab = 0$, the effective coupling between $\state{i}$ and
$\state{f}$ is so small that it is lost in the course of
the approximation.
Hence, this approximation provides a useful hint for the least favorable
combination of parameters which results in the weakest effective
coupling between $\state{i}$ and $\state{f}$; consequently, the
adiabatic limit is approached most slowly in this case.

The adiabatic-elimination approximation allows to estimate
how quickly the adiabatic limit is approached when the AT state exists.
Then, as we noted above, we have a level-crossing transition,
the probability for which can be roughly described
by the Landau-Zener formula,
\be{LZformula}
P_f \approx 1-e^{-\pi\Omeff^2(t_c)/\dot\Deff(t_c)},
\ee
where $t_c$ is the crossing point.
For the Gaussian shapes (\ref{shapes})
we have $t_c=(T^2/8\tau)\ln(\Sb/\Sa)$ and
\be{LZparam-full}
\frac{\Omeff^2(t_c)}{\dot\Deff(t_c)} = (\Om_0 T)^2\LZ,
\ee
with
\be{LZparam}
\LZ = \frac{T}{4\tau} \frac{\Sab^2}{\sqrt{\Sa\Sb}}
	\exp\left[-\frac{2\tau^2}{T^2}
	  -\frac{T^2}{32\tau^2}\left(\ln\frac\Sb\Sa\right)^2\right].
\ee
The larger this parameter, the faster the adiabatic limit is approached.
We thus conclude that from the adiabaticity viewpoint, the most favorable
case is when $\Sa=\Sb$ and the ratio $\Sab^2/\sqrt{\Sa\Sb}$ is large.
Not surprisingly, the Landau-Zener parameter (\ref{LZparam-full}) is
also proportional to $(\Om_0 T)^2$, which is essentially the squared
pulse area.


\begin{figure}[tbp]
\centerline{\psfig{width=75mm,file=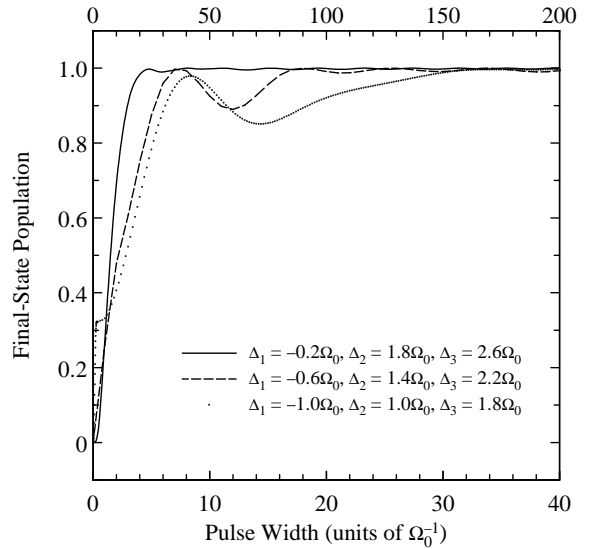}}
\vspace*{3mm}
\caption{
The final-state population $P_f$ as a function of the pulse width $T$ in
the case of $N=3$ intermediate states for three combinations of
detunings, denoted in the figure.
The upper $T$-scale is for the dotted curve while
the lower $T$-scale is for the dashed and solid curves. 
The coupling strengths are the same in all cases,
$\a_1=\b_1=1$, $\a_2=0.6$, $\b_2=1.0$, $\a_3=1.2$, $\b_3=0.6$.
}
\label{Fig-Sab}
\end{figure}


In Fig.~\ref{Fig-Sab} the final-state population $P_f$ is plotted
against the pulse width $T$ in the case of $N=3$ intermediate states
for three combinations of detunings.
The coupling strengths are the same in all cases.
The parameters for the dotted curve are chosen so that $\Sab=0$
while those for the other curves ensure that $\Sab\neq 0$.
The parameter (\ref{LZparam}) is
$\LZ=0$ for the dotted curve,
$\LZ\approx 0.326$ for the dashed curve, and
$\LZ\approx 1.367$ for the solid curve.
As a result, the adiabatic limit is approached
most slowly for the dotted curve,
more quickly for the dashed curve,
and most quickly for the solid curve.


\subsection{The on-resonance case}

When a certain intermediate state $\state{n}$ is on single-photon resonance,
$\Delta_n=0$, it cannot be eliminated adiabatically.
By eliminating all other intermediate states, the general $N+2$-
state problem is reduced to an effective three-state problem,
\be{AE3}
i \frac{d}{dt} \bM{ c_i \cr c_n \cr c_f }\eM \approx
\bM{
\OmP^2\San    \& \a_n\OmP \& \OmP\OmS\Sabn \cr
\a_n\OmP      \& 0        \& \b_n\OmS \cr
\OmP\OmS\Sabn \& \b_n\OmS \& \OmS^2\Sbn
}\eM
\bM{ c_i \cr c_n \cr c_f }\eM,
\ee
where the $\S^{(n)}$ sums are defined by Eqs.~(\ref{sums-n}).
Comparison with the standard three-state STIRAP shows that the
off-resonant states induce dynamic Stark shifts of states $\state{i}$
and $\state{f}$, which result in a nonzero two-photon detuning.
Moreover, the off-resonant states induce a direct coupling between
states $\state{i}$ and $\state{f}$.
Careful examination of Eq.~(\ref{AE3}) shows
that the AT state (\ref{AT}) always exists, but
it involves in general a nonzero contribution
from the intermediate state $\state{n}$.
This contribution vanishes when the proportionality condition
(\ref{proportional}) is fulfilled; then $\San=\Sbn=\Sabn$ and the AT state
is the dark state, $\AT(t)\equiv \dark(t)$.
In particular, if $\San=\Sbn=\Sabn=0$, the multi-$\Lambda$ system
behaves exactly like STIRAP.


\section{Discussion and conclusions}

\label{Sec-conclusion}

We have presented an analytic study, supported by numerical examples,
of adiabatic population transfer from an initial state $\state{i}$
to a final state $\state{f}$ via $N$ intermediate states by means of
two delayed and counterintuitively ordered laser pulses.
Thus this paper generalizes the original STIRAP,
operating in a single three-state $\Lambda$-system,
to a multistate system involving $N$ parallel $\Lambda$-transitions.
The analysis has shown that the dark state $\dark(t)$,
which is a linear combination of $\state{i}$ and $\state{f}$ and
transfers the population between them in STIRAP,
remains a zero-eigenvalue eigenstate of the Hamiltonian (\ref{H})
only when condition (\ref{proportional}) is fulfilled.
Hence, in this case the multi-$\Lambda$ system behaves very similarly
to the single $\Lambda$-system in STIRAP.
Condition (\ref{proportional}), which is essentially a relation
between the transition dipole moments, requires that for each
intermediate state $\state{k}$, the ratio $\Om_{P,k}(t)/\Om_{S,k}(t)$
between the couplings to states $\state{i}$ and $\state{f}$
is the same and does not depend on $k$.
Moreover, this condition ensures the existence of the dark state both
in the case when all intermediate states are off single-photon resonance
and when one or more states are on resonance.

When condition (\ref{proportional}) is not fulfilled the dark state
$\dark(t)$ does not exist but a more general adiabatic-transfer state
$\AT(t)$, which links adiabatically the initial and final states
$\state{i}$ and $\state{f}$, may exist under certain conditions.
Unlike $\dark(t)$, state $\AT(t)$ contains contributions from
the intermediate states which therefore acquire transient populations
during the transfer.
We have shown that when one and only one intermediate state
is on resonance, the AT state always exists.
When more than one intermediate states are on resonance, the AT state
exists only when the proportionality relation (\ref{proportional}) is
fulfilled, at least for the degenerate states.
In the off-resonance case, the condition for existence of $\AT(t)$
is given by Eq.~(\ref{ATcondition}) which is a condition on
the single-photon detunings and the relative coupling strengths.
It follows from this condition that when the pump and Stokes
frequencies are scanned across the intermediate states
(while maintaining the two-photon resonance),
the final-state population $P_f$ passes through $N$ regions of high
transfer efficiency (unity in the adiabatic limit) and $N-1$ regions of
low efficiency (zero in the adiabatic limit).
Each of the low-efficiency regions is situated between two
adjacent intermediate states,
while each intermediate state is within a region of high efficiency,
as shown in Figs.~\ref{Fig-2S-D} and \ref{Fig-5S-D}.
Our results suggest that it is most appropriate to tune the pump and
Stokes lasers either just below or just above all intermediate states
because there
firstly, the AT state always exists;
secondly, the adiabatic regime is achieved more quickly;
thirdly, the transfer is more robust
 against variations in the laser parameters;
and fourthly, the transient intermedaite-state populations,
which are proportional to $\Delta_k^{-2}$, can easily be suppressed.


\section*{Acknowledgments}

This work has been supported financially by the Academy of Finland.


\end{document}